\documentclass[11pt]{article}

\usepackage{authblk}
\usepackage[utf8]{inputenc}
\usepackage{amsfonts,amsmath,bm,bbm}
\usepackage{graphicx}
\usepackage{epstopdf}
\usepackage{amssymb}
\usepackage{hyperref}
\usepackage{natbib}
\usepackage[table,xcdraw]{xcolor}
\usepackage{array,multirow,graphicx}

\usepackage[a4paper,margin={2.5cm,3.5cm}]{geometry}

\DeclareMathOperator{\argmin}{argmin}

\usepackage{color}
\definecolor{gray}{rgb}{0.5,0.5,0.5}
\definecolor{dgreen}{rgb}{0,0.5,0}

\begin{document}

\title{\textbf{Forecasting Electricity Prices}}

\author{\small Katarzyna Maciejowska, Bartosz Uniejewski, Rafa{\l} Weron}

\affil{\small \textit{Department of Operations Research and Business Intelligence,\linebreak Wroc{\l}aw University of Science and Technology, 50-370 Wroc{\l}aw, Poland}}

\date{\small This version: \today}

\maketitle

\begin{abstract}
Forecasting electricity prices is a challenging task and an active area of research since the 1990s and the deregulation of the traditionally monopolistic and government-controlled power sectors. It is interdisciplinary by nature. It requires expertise in econometrics, statistics or machine learning for developing well-performing predictive models, in finance for understanding market mechanics, and in electrical engineering for comprehension of the fundamentals driving electricity prices. 

Although electricity price forecasting aims at predicting both spot and forward prices, the vast majority of research is focused on short-term horizons which exhibit dynamics unlike in any other market. The reason is that power system stability calls for a constant balance between production and consumption, while being weather (both demand and supply) and business activity (demand only) dependent. The recent market innovations do not help in this respect. The rapid expansion of intermittent renewable energy sources is not offset by the costly increase of electricity storage capacities and modernization of the grid infrastructure.

On the methodological side, this leads to three visible trends in electricity price forecasting research as of 2022. Firstly, there is a slow, but more noticeable with every year, tendency to consider not only point but also probabilistic (interval, density) or even path (also called ensemble) forecasts. Secondly, there is a clear shift from the relatively parsimonious econometric (or statistical) models towards more complex and harder to comprehend, but more versatile and eventually more accurate statistical/machine learning approaches.  Thirdly, statistical error measures are nowadays regarded as only the first evaluation step. Since they may not necessarily reflect the economic value of reducing prediction errors, more and more often, they are complemented by case studies comparing profits from scheduling or trading strategies based on price forecasts obtained from different models. 


\vspace*{.5cm}

\end{abstract}

\section{Introduction}

\textit{Electricity price forecasting} (EPF)\footnote{We use EPF when referring to both electricity price forecasting and
electricity price forecast(s). The plural form, i.e., forecasts, is abbreviated EPFs.} as a research area of its own appeared in the early 1990s with the liberalization and deregulation of the power sectors in the UK and Scandinavia. The late 1990s and 2000s were marked by the widespread conversion from the traditionally monopolistic and government-controlled power sectors to competitive power markets in Europe, North America, Australia and eventually in Asia \citep{may:tru:18}. Over the years, EPFs have become a fundamental input to companies' decision-making mechanisms \citep{wer:14}. As \cite{hon:15} estimates, for a medium-sized utility with a 5-gigawatt annual peak load\footnote{The annual peak load is the highest electrical power demand in a (calendar) year. The power consumed or generated is measured in multiples of the watt (W). Smaller power plants can generate tens of megawatts (1~MW = $10^6$~W), the largest tens of gigawatts (1~GW = $10^9$~W). The amount of electricity consumed or generated over a specific period of time is typically measured in megawatt-hours (MWh); it is also the basic unit used in trading electricity.}, improving the day-ahead demand forecasts by 1\% leads to annual savings of ca. 1.5 million USD. With the additional price forecasts, the savings double. Clearly, the time invested in developing EPF models can pay off. 

For newcomers to this research area it is important to realize that the literature has generally focused on horizons of up to 48 hours, since short-term price dynamics is what makes electricity special. In the longer term, prices are averaged across weekly, monthly or annual delivery periods and lose much of their uniqueness. In the short-term, on the other hand, electricity prices exhibit significant seasonality at different levels (daily, weekly and in many markets also annual), short-lived and generally unanticipated price spikes (ranging up to two orders of magnitude), and in some markets even negative values. 

However, the ``short-term'' is not a particular horizon, but a whole spectrum of horizons ranging from a few minutes ahead (real-time, \textit{intraday}, ID; also called ``spot'' in North America) to a \textit{day-ahead} (DA; called ``spot'' in Europe). Note that from a financial perspective, both the ID and DA contracts can be regarded as very short-term forwards, with delivery during a particular hourly (half- or quarter-hourly) load period on the same or the next day. 
Since each day can be divided into a finite number of load periods $h=1,2,...,H$ with $H=24, 48$ or $96$, it is common to use double indexing when referring to the electricity price. Here, we denote by $P_{d,h}$ the price for day $d$ and load period $h$, by $\widehat{P}_{d,h}$ its point forecast and by $\widehat{F}_{P}$ or $\widehat{F}_{P_{d,h}}$ its predictive distribution. 

In what follows, we provide an overview of EPF research, with a particular focus on three current trends: 
\begin{description}
    \item[~~\#1:] increasing popularity of \textit{probabilistic} (interval, density) and \textit{path} (also called \textit{ensemble}) forecasts,
    \item[~~\#2:] a visible shift towards \textit{statistical/machine learning} (SL/ML), and 
    \item[~~\#3:] evaluating the economic value of price predictions. 
\end{description}
However, before we start, let us first briefly describe the marketplace and the typical forecasting tasks considered.

\section{The Marketplace}

As a result of the aforementioned liberalization and deregulation of the power sectors, two basic models for power markets have emerged: power pools -- where trading, dispatch and transmission are managed by the \textit{system operator} (SO), and power exchanges -- where trading and initial dispatch are managed by an institution independent from the \textit{transmission system operator} (TSO). Participation in power pools is limited to generators and is typically mandatory. The \textit{market clearing price} (MCP) is established through a one-sided auction as the intersection of the supply curve constructed from aggregated supply bids of the generators and the demand predicted by the system operator. Often a separate price for each node in the network is calculated, so-called \textit{locational marginal price} (LMP). Such a system was adopted in highly meshed North American networks. On the other hand, in Australia, where the network structure is simpler, zonal pricing was successfully implemented, where for areas without grid limitations a unique price is settled.

\begin{figure}[tb]
	\centering
    \includegraphics[width=1\textwidth]{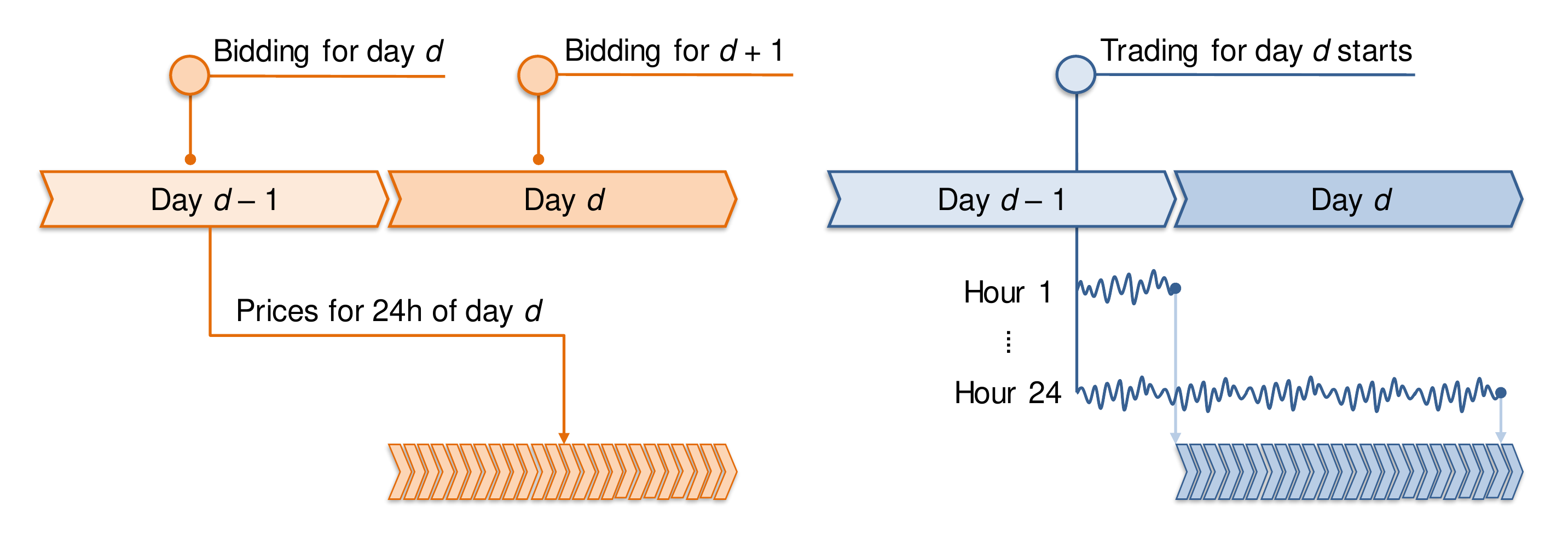}
    \caption{Illustration of bidding and price settlement in auction (\textit{left}) and continuous trading (\textit{right}) power markets. In day-ahead auctions the bids for all load periods (here: hours) of day $d$ can be submitted until a certain hour on day $d-1$. Intraday markets which admit continuous trading run 24/7 from an afternoon hour on day $d-1$ up until a few minutes before the delivery on day $d$.
    } 
	\label{fig:DA_ID_bidding}
\end{figure}

In contrast to power pools, participation in power exchanges is -- except for some special cases -- voluntary and open not only to generators, but also to wholesale consumers and speculators. The price is established either through a two-sided auction (DA, ID) as the intersection of the supply curve constructed from aggregated supply bids and the demand curve constructed from aggregated demand bids or in continuous trading (ID). Most market designs have adopted the uniform-price auction, where buyers who bid at or above the MCP pay that price and sellers who bid at or below the MCP are paid this price. Moreover, in auction markets the bids can be submitted until a certain time -- called \textit{gate closure} -- which is the same for all load periods, see the left panel in Fig. \ref{fig:DA_ID_bidding}. Hence, auction prices could be viewed as realizations of a multivariate random variable and therefore prices for all load periods should be predicted simultaneously \citep{zie:wer:18}. On the other hand, some ID markets allow for continuous trading. They run 24/7 from an afternoon hour on day $d-1$ up until a few minutes before the delivery of electricity during a particular load period on day $d$, see the right panel in Fig. \ref{fig:DA_ID_bidding}.

In some countries (e.g., Germany, Ireland, Poland) the DA and ID markets are complemented by the so-called balancing market. This technical market is used for pricing differences between the market schedule and actual system demand for very short time horizons before delivery. For instance, the TSO might instruct a generator to increase its output to meet a sudden surge in demand. The producer then receives a premium via the balancing market for the energy generated used to balance the grid. 

\begin{figure}[tb]
	\centering
    \includegraphics[width=1\textwidth]{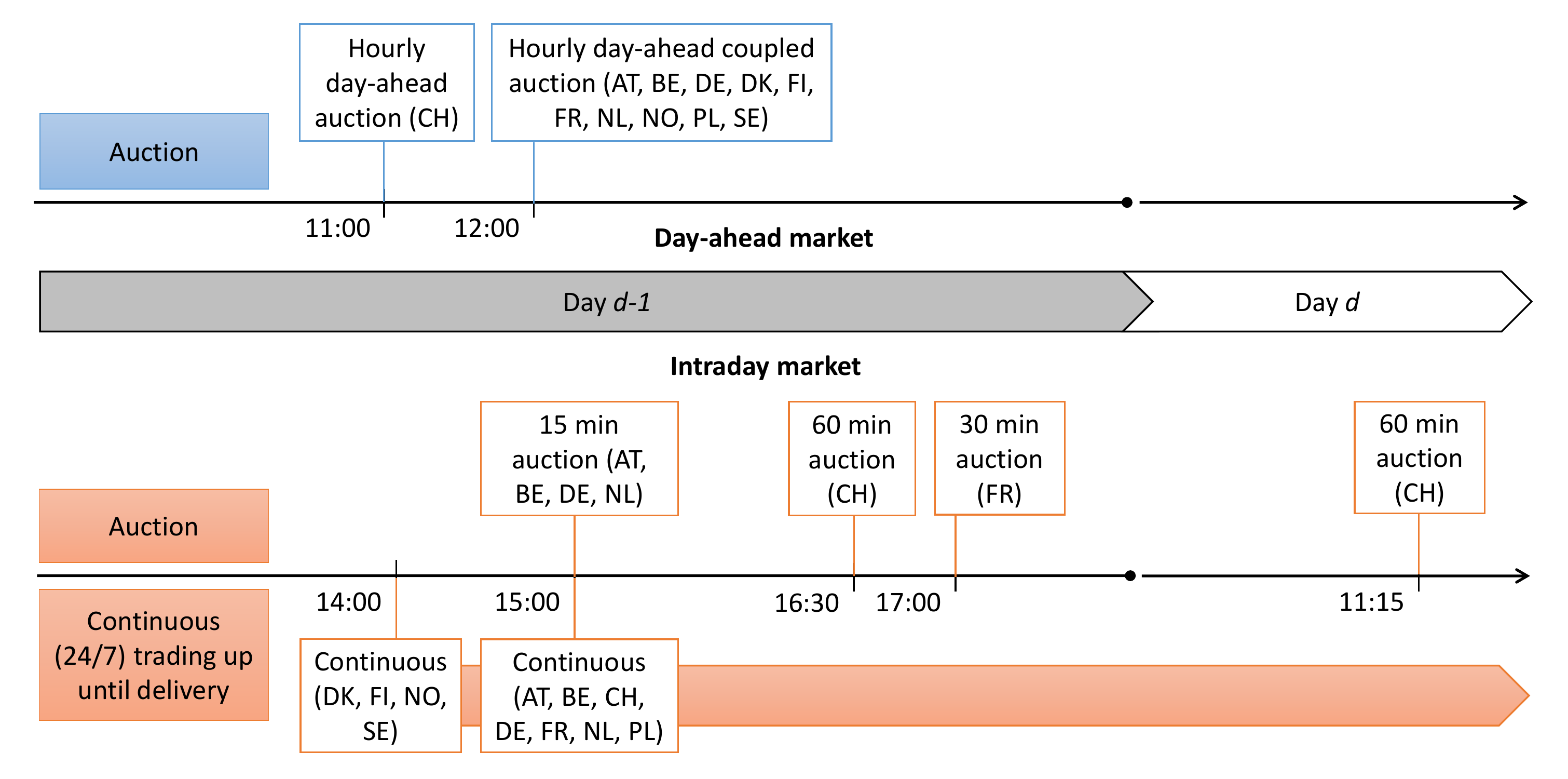}
    \caption{The timeline of day-ahead (\textit{top}) and intraday (\textit{bottom}) trading activities for delivery of electricity on day $d$ in selected European countries: Austria (AT), Belgium (BE), Denmark (DK), Germany (DE), Finland (FI), France (FR), the Netherlands (NL), Norway (NO), Poland (PL), Sweden (SE) and Switzerland (CH). 
    } 
	\label{fig:timeline}
\end{figure}

The timeline of day-ahead and intraday trading activities in selected European countries is illustrated in Fig. \ref{fig:timeline}. As can be seen, the DA and ID markets complement each other. Once the gate closes for day-ahead bids around noon, various intraday markets open for adjusting these bids. They are particularly important for nondispatchable, stochastic producers such as wind or solar farms, and include both auctions and continuous trading. Note that both the ID and DA contracts can concern delivery during the same load period, only the time the decision has to be made and the bid placed differs. 

The presented sequence of events has important implications for study design. In the DA market the forecasting horizons typically range from 12-14 hours for the first load period of the next day to 36-38 hours for the last. However, at the time the predictions are made, i.e., the morning hours of day $d-1$, the DA prices for all load periods of this day are already known (they were settled around noon on day $d-2$). Generally, the TSO day-ahead forecasts of the system load ($\approx$ demand) and the system-wide generation from \textit{renewable energy sources} (RES) are also available to market participants at this time. 

When the ID market is considered, the selection of the forecasting horizon depends on the research question. Firstly, the predictions can be made on the morning of day $d-1$, when  market participants need to decide how much electricity to bid in the DA market and how much to buy/sell in the ID market or leave for the balancing market. Forecasts of the price spread between DA and ID/balancing markets can provide valuable insights for decision-making \citep{mac:nit:wer:19,mac:nit:wer:21}. 

Secondly, the predictions can be used for bidding in ID markets with continuous trading. Although the trading floor opens in the afternoon hours of day $d-1$, the majority of bids are placed during the last 3-4 hours before the delivery \citep{nar:zie:19}. Hence, the forecasting horizons considered typically range from a couple of minutes to 4 hours \citep{jan:ste:19,uni:mar:wer:19:IJF:IDlasso,nar:zie:20b}. Note that different model specifications may be optimal for predicting ID prices for different horizons \citep{mac:uni:ser:20}.  
Since the bidding behavior of market participants is significantly influenced by RES generation forecasts which are available at the time of trading \citep{kie:par:17,kul:zie:21}, ID price forecasts should not only exploit the short-term price dependencies but also updated predictions of wind and solar power generation. Interestingly, including self-exciting terms in ID models allows to better capture the empirically observed trade clustering \citep{kra:kie:21}.

\section{Trend \#1: From Point to Probabilistic and Ensemble Forecasts}
\label{sec:Trend1}

By far point forecasts are the most popular. Despite a few early attempts, often inspired by developments in wind forecasting \citep{hon:pin:etal:20}, probabilistic forecasting was not part of the mainstream EPF literature until 2014 and the Global Energy Forecasting Competition \cite[GEFCom2014;][]{hon:pin:fan:etal:16}. Probabilistic EPF quickly gained momentum and energy analysts have become aware of its importance in energy systems planning and operations. A variety of approaches have been considered, including bootstrapping \citep{che:etal:12}, QRA (see Section \ref{ssec:QRA}), Bayesian statistics \citep{kos:kos:19} and deep learning \citep{mas:etal:21,jed:lag:mar:wer:22}. 
Nevertheless, as of today, no more than 15\% of Scopus-indexed articles concern interval or distributional EPF. 
Path (also called \textit{ensemble}) forecasts, which focus on the multidimensional temporal distribution, are even less popular. Yet, path-dependency is crucial for many optimization problems arising in power plant scheduling, energy storage and trading, and this has been recognized in the recent EPF literature \citep{jan:ste:20,nar:zie:20b}.

\subsection{Error and Price Distributions}

There are two main approaches to probabilistic forecasting: the more elegant one directly considers the distribution of the electricity price, while the more popular one builds on the point forecast and the distribution of errors associated with it. In both cases, the focus can be on prediction intervals, selected quantiles or the whole predictive distribution. For reviews on short- and medium-term probabilistic EPF we refer to \cite{now:wer:18} and \cite{zie:ste:18}, respectively, while for a general treatment to the  seminal review of \cite{gne:kat:14}.

If we assume that the point forecast is the expected price\footnote{Although this is the most common assumption, the point forecast does not have to be the expected value. For instance, it can be the median or any quantile of the predictive distribution.} at a future time point, i.e, $\widehat{P}_{d,h} = \mathbb{E}(P_{d,h})$, then we have: 
\begin{equation}
    P_{d,h} = \widehat{P}_{d,h} + \varepsilon_{d,h},
\end{equation}
and the distribution $F_{\varepsilon}$ of errors associated with $\widehat{P}_{d,h}$ is identical to the distribution $F_P$ of prices, except for a horizontal shift by $\widehat{P}_{d,h}$:
\begin{equation*}
    F_{P}(x) \equiv \mathbb{P}(P_{d,h} \leq x) = \mathbb{P}(\widehat{P}_{d,h} + \varepsilon_{d,h} \leq x) = \mathbb{P}(\varepsilon_{d,h} \leq x - \widehat{P}_{d,h}) \equiv F_{\varepsilon}(x - \widehat{P}_{d,h}).
\end{equation*}
This, however, implies that the inverse empirical \emph{cumulative distribution functions} (also called \textit{quantile functions}) satisfy:
\begin{equation}
\label{eqn:PriceAsForecastPlusError:InvECDFs}
\widehat{F}^{-1}_{P}(\alpha) = \widehat{P}_{d,h} + \widehat{F}^{-1}_{\varepsilon}(\alpha),
\end{equation}
i.e., they are identical except for a shift by $\widehat{P}_{d,h}$, but this time on the vertical axis. Equation \eqref{eqn:PriceAsForecastPlusError:InvECDFs} provides the basic framework for constructing probabilistic forecasts from prediction errors. If a dense grid of quantiles is considered, e.g., 99 percentiles, then $\widehat{F}_{P}$ can be approximated pretty well \citep{hon:pin:fan:etal:16,now:wer:18,uni:wer:21}.

If we assume that $F_{P}$ has a density $f_{P}$, then a density forecast $\widehat{f}_{P}$ can be provided as well. However, \cite{zie:ste:16} argue against using such an approach. Analyzing the fine structure of aggregated supply and demand curves in the German market they found that $F_P$ was multimodal with significant jumps (corresponding to point masses) at certain `round' prices.

\subsection{Quantile Regression Averaging}
\label{ssec:QRA}

\textit{Quantile regression} \cite[QR; see][]{koe:05} is one of the most popular methods for directly modeling the distribution of a random variable. QR approximates the target quantile with a linear function of a set of explanatory variables. In the EPF context, these variables typically contain publicly available market information \cite[load forecasts, generation structure,  historical electricity prices, etc.;][]{bun:and:che:wes:16,mac:20} and/or point predictions of electricity prices \citep{wer:14}. The later case leads to the so-called \textit{Quantile Regression Averaging} (QRA) introduced by \cite{now:wer:15} and originally developed for Team Poland's participation in the GEFCom2014 competition \citep{mac:now:16}. It is a forecast combination approach to the computation of quantile forecasts, which bridges the gap between point and probabilistic forecasts. QRA involves applying QR to the point forecasts of a small number of individual forecasting models or experts:
\begin{equation}\label{eq:qr}
q(\alpha|\boldsymbol{X}_{d,h}) = \boldsymbol{X}_{d,h} \boldsymbol\beta_\alpha,
\end{equation}
where $q(\alpha | \cdot)$ is the conditional $\alpha$th quantile of $F_P$, $\boldsymbol{X}_{d,h}$ is the vector of point forecasts and $\boldsymbol\beta_{\alpha}$ is the corresponding vector of weights.
The latter is estimated by minimizing the following sum of \textit{check functions}: 
\begin{equation}\label{eq:qr:loss}
	\widehat{\boldsymbol\beta}_\alpha
	= \underset{\boldsymbol\beta_{\alpha}}\argmin \Big\{ \textstyle\sum_{d,h} \underbrace{\left({\alpha}-\mathbbm{1}_{{P}_{d,h}< {\boldsymbol{X}_{d,h}}\boldsymbol\beta_{\alpha}}\right)\left({P}_{d,h} - \boldsymbol{X}_{d,h} \boldsymbol\beta_\alpha\right)}_{\mbox{\scriptsize {check function}}} \Big\}.
\end{equation}
The very good forecasting performance of QRA has been verified by a number of authors, not only in the area of EPF \citep{liu:now:hon:wer:17,kos:kos:19,kat:zie:21,uni:wer:21}. However, its most spectacular success came during the GEFCom2014 competition, when teams using variants of QRA \citep{gai:gou:ned:16,mac:now:16} were ranked in the top two places in the price track.

\subsection{Paths and Ensembles}

Although the concept of probabilistic EPF is much more general than of point forecasting, it is not sufficient to support operational decisions that depend on future trajectories of electricity prices. For instance, in Germany renewable energy producers can receive less subsidies if the electricity price is negative for 6 hours in a row.
Hence, instead of looking at the 24 hourly univariate price distributions $F_{P_{d,1}},...,F_{P_{d,24}}$, we should be focusing on the multidimensional distribution $\boldsymbol{F}_{\boldsymbol{P}}$ of the 24-dimensional price vector $\boldsymbol{P}_{\boldsymbol{d}} = (P_{d,1},\ldots,P_{d,24})'$. However, many models considered in the literature cannot output such a multidimensional forecast.

\emph{Ensemble} forecasts provide a practical remedy. An ensemble is a collection of simulated price \textit{paths}, also called \textit{trajectories} or \textit{scenarios}. For a large number of paths the ensemble approximates the underlying distribution ${\boldsymbol{F}}_{\boldsymbol{P}}$ arbitrarily well \citep{wer:zie:20}. In practice `large' means thousands or millions of paths, which may be a computational challenge \citep{nar:zie:20b}. It should be noted that, on one hand, the same or similar concepts have been used in different disciplines under different names, e.g., \emph{simultaneous prediction intervals}, \emph{prediction bands}, \emph{spatio-temporal trajectories}, \emph{numerical weather prediction ensembles}. On the other, the term ensemble is also used to refer to any averaging of -- point or probabilistic -- forecasts \citep{hon:pin:etal:20}.

\section{Trend \#2: From Regression to Statistical and Machine Learning}
\label{sec:Trend2}

Until the mid 2010s, the EPF literature was dominated by relatively parsimonious linear regression and neural network models. They were characterized by a small number -- a dozen or two -- of \textit{inputs} (also called \textit{features}, \textit{input features}, \textit{explanatory variables}, \textit{regressors}, or \textit{predictors}) and complex data pre-/post-processing:
\begin{itemize}
    \item replacing outliers, i.e., price spikes, by more `normal' values before estimating the model \citep{con:esp:nog:con:03,bie:men:rac:tru:07,jan:tru:wer:wol:13} or utilizing robust estimation methods \citep{gro:nan:19};

    \item averaging forecasts, both across models \citep{nan:09,bor:bun:lis:nan:13,now:rav:tru:wer:14} and across calibration windows for the same model \citep{mar:ser:wer:18,hub:mar:wer:19}, also in a probabilistic EPF setting \cite[see the QRA approach in Section \ref{ssec:QRA} and][]{ser:uni:wer:19};
    
    \item using so-called variance stabilizing transformations \cite[VSTs;][]{sch:11,dia:pla:16,uni:wer:zie:18,nar:zie:19,shi:wan:che:ma:21} to make the marginal distributions less heavy-tailed (Box-Cox family, area hyperbolic sine) or Gaussian (Probability Integral Transform, see Section \ref{sssec:PEPF});
    
    \item deseazonalizing the data with respect to the long-term seasonal component (LTSC) before estimating the model \citep{jan:tru:wer:wol:13,now:wer:16,lis:pel:18,afa:fed:19,uni:mar:wer:19ENEECO,mar:uni:wer:20IJF}. 
\end{itemize}

However, as more data and computational power became available, the models became more complex to the extent that expert knowledge was no longer enough to handle them \citep{jed:lag:mar:wer:22}. This paved the way for statistical/machine learning in EPF.
Arguably, \textit{statistical learning} (SL) and \textit{machine learning} (ML) are synonyms.\footnote{\cite{jan:etal:20} even argue that the distinction between \textit{statistical} and \textit{machine learning} forecasting is dubious, as this distinction does not stem from fundamental differences in the methods assigned to either class, but rather is of a ``tribal'' nature.} They have just originated in different communities -- computer science/artificial intelligence \citep{mit:97} or computational statistics \citep{jam:wit:has:tib:21}. Both SL and ML refer to a vast set of (computational, statistical) tools for understanding data, both can improve ``automatically'' through training. In either case, learning can be supervised or unsupervised. In EPF we are typically interested in supervised learning, which involves building a model for predicting a known output or outputs based on a set of inputs.

\subsection{The Expert Benchmark}
\label{ssec:expert}

A class of commonly used EPF benchmarks is based on a parsimonious \textit{autoregressive} (AR) structure with exogenous variables and calendar effects, originally proposed by \cite{mis:tru:wer:06}. Since expert knowledge is used to select the regressors, such benchmarks are often called \textit{expert} models \citep{zie:wer:18}. One of the most popular structures represents the electricity price for day $d$ and hour $h$ by:
\begin{eqnarray}
P_{d,h} & = & \underbrace{\beta_1 P_{d-1,h} + \beta_2 P_{d-2,h} + \beta_3 P_{d-7,h}}_{\mbox{\scriptsize autoregressive effects}} + \underbrace{\beta_4 P_{d-1,24}}_{\mbox{\scriptsize end-of-day}} + \underbrace{\beta_5 P_{d-1}^{max} + \beta_6 P_{d-1}^{min}}_{\mbox{\scriptsize non-linear effects}} \nonumber \\
&& + \underbrace{\beta_7 X^1_{d,h} + \beta_8 X^2_{d,h}}_{\mbox{\scriptsize exogenous variables}} + \underbrace{\sum\nolimits_{j=1}^7 \beta_{h,j+8} D_{j}}_{\text{weekday dummies}} + \varepsilon_{d,h}, 
\label{eqn:expert}
\end{eqnarray}
where $P_{d-1,h}$, $P_{d-2,h}$ and $P_{d-7,h}$ account for the autoregressive effects and correspond to prices from the same hour $h$ of the previous day, two days before and a week before, $P_{d-1,24}$ is the last known price at the time the prediction is made and provides information about the end-of-day price level, $P^{max}_{d-1}$ and $P^{min}_{d-1}$ represent previous day's maximum and minimum prices,  $X^1_{d,h}$ and $X^2_{d,h}$ are exogenous variables, $D_{1},..., D_{7}$ are weekday dummies and $\varepsilon_{d,h}$ is the noise term (i.i.d.\ variables with finite variance). The $\beta_i$'s are estimated using \textit{ordinary least squares} (OLS).

\begin{figure}[tb]
	\centering
    \includegraphics[height=5cm]{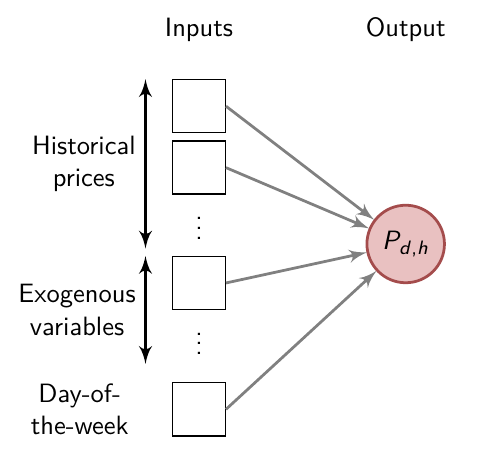}~~~~~~
    \includegraphics[height=5cm]{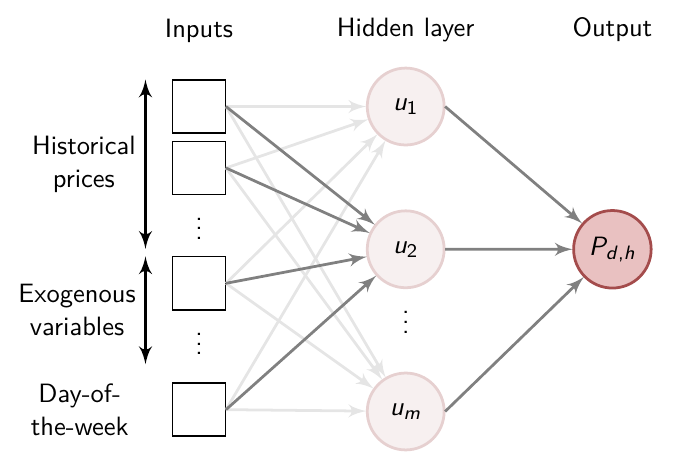}
    \caption{Visualization of a linear regression model (\textit{left}) and a shallow neural network (\textit{right}) with identical inputs and output, i.e., the electricity price for day $d$ and hour $h$. White squares represent the inputs and $u_1, u_2, ..., u_m$ the hidden nodes (or neurons). Arrows indicate the flow of information. 
    } 
	\label{fig:ARX:NN}
\end{figure}

Autoregression or more generally linear regression is one of the two most commonly used classes of EPF models \citep{wer:14}. The other is the \textit{multi-layer perceptron} (MLP). The simplest neural network, a single-layer perceptron, contains no hidden layers (only inputs and the output) and is equivalent to a linear regression -- both represent $P_{d,h}$ by a linear combination of  input features, see the left panel in Fig.\ \ref{fig:ARX:NN}. On the other hand, the MLP includes at least one hidden layer and utilizes a feed-forward architecture -- the outputs of the nodes (or neurons) in one layer are inputs to the next one, see the right panel in Fig.\ \ref{fig:ARX:NN}. Since the output of a node is a weighted sum of all of the inputs transformed by a typically nonlinear activation function, unlike in linear regression, NNs can tackle complex dependence structures encountered in power market data \citep{kel:etal:16}. 

Exogenous variables typically include day-ahead predictions of the system load and RES generation \citep{lag:mar:sch:wer:21}. Days with high demand and low RES generation are characterized by relatively high prices. On the other hand, high RES generation pulls prices down; in periods of low demand -- holidays and/or at night -- even below zero \citep{zho:s-w:sec:smi:16}. 
Other fundamental variables may possess explanatory power as well. For instance, fuel and CO$_2$ allowance prices, especially for medium-term EPF \citep{mac:wer:16, zie:ste:18}. Due to the merit order effect, i.e., dispatching units characterized by the lowest marginal cost of production, the fuel--electricity price relationship changes throughout the day. Natural gas prices impact mainly the peak hours, whereas coal prices influence the off-peak hours. 
Finally, the day-of-the-week input feature visible in Fig.\ \ref{fig:ARX:NN} can be a set of weekday dummies, as in Eqn.\ \eqref{eqn:expert}, or a single multi-valued variable, which is more common in NN models.

\subsection{Regularization and the LEAR Model}
\label{ssec:LEAR}

Selecting regressors is a cumbersome task and expert knowledge does not always identify the relevant ones. In a series of papers in the mid 2010s, \cite{lud:feu:neu:15}, \cite{zie:ste:hus:15}, \cite{gai:gou:ned:16}, \cite{uni:now:wer:16} and \cite{zie:16:TPWRS} introduced the concept of regularization to EPF. In simple terms, the idea behind this approach is to add a penalty term to the \textit{residual sum of squares} (RSS) in OLS regression:
\begin{equation}
\label{eq:LASSO}
\widehat{\boldsymbol\beta} = \underset{\boldsymbol\beta}\argmin \left\{
\text{RSS}
+ \lambda \sum_{i=1}^{n} |\beta_{i}|^q\right\},
\end{equation}
where $\lambda$ is the \emph{tuning} or \emph{regularization} hyperparameter. Note, that \textit{hyperparameters} are model parameters that cannot be optimized during the  training (estimation) phase, but have to be set or calibrated beforehand, e.g., using cross-validation \citep{jam:wit:has:tib:21}. For $q=2$ we obtain \textit{ridge regression} \citep{hoe:ken:70} and for $q=1$ the \emph{least absolute shrinkage and selection operator} \cite[LASSO; ][]{tib:96}. The latter can shrink $\beta_i$'s not only towards zero but actually to zero itself, thus effectively eliminating some regressors from the model. If we admit both terms in Eqn.\ \eqref{eq:LASSO}, i.e., $\lambda_1 \sum |\beta_i| + \lambda_2 \sum \beta_i^2$, then we obtain the so-called \textit{elastic net}  \citep{zou:has:05}. 


In the EPF setting, all three variants were compared in \cite{uni:now:wer:16}. Ridge regression easily outperformed expert models and stepwise regression techniques, but was significantly worse than the LASSO and the elastic net. At the cost of an additional parameter, the elastic net generally yields more accurate predictions than the LASSO. Nevertheless, the latter has become the golden standard in EPF \citep{uni:wer:18,zie:wer:18,jan:ste:19,nar:zie:19,mar:20,zha:li:ma:20,oze:yil:21}. 
It was even utilized by \cite{lag:mar:sch:wer:21} to construct a well-performing EPF benchmark -- the \textit{LASSO-Estimated AutoRegressive} (LEAR) model. The starting point for the LEAR model is a parameter-rich regression:
\begin{equation}\label{eq:LEAR}
\begin{aligned}
P_{d,h} &= \sum_{i=1}^{24} \left( \beta_{h,i} P_{d-1,i} 
+ \beta_{h,i+24} P_{d-2,i} + \beta_{h,i+48} P_{d-3,i}
+ \beta_{h,i+72}P_{d-7,i} \right)  \\
&\quad + \sum_{i=1}^{24} \left( \beta_{h,i+96} X^1_{d,i} + \beta_{h,i+120} X^1_{d-1,i} + \beta_{h,i+144} X^1_{d-7,i} \right)  \\
&\quad + \sum_{i=1}^{24} \left( \beta_{h,i+168} X^2_{d,i} + \beta_{h,i+192} X^2_{d-1,i} + \beta_{h,i+216} X^2_{d-7,i} \right)  \\
&\quad + \sum_{k=1}^{7} \beta_{h,240+k} D_k  +\varepsilon_{d,h},
\end{aligned}
\end{equation}
which differs from the expert model in Eqn.\ \eqref{eqn:expert} mainly by allowing for cross-hourly dependencies. In general, the price for hour $h$ may depend on the prices for all 24 hours yesterday, the day before, etc. In practice, only a dozen or two of the potential 247 regressors turn out to be relevant. However, they need not be the ones included in the expert model.

\begin{figure}[tb]
\centering
    \includegraphics[height=7cm]{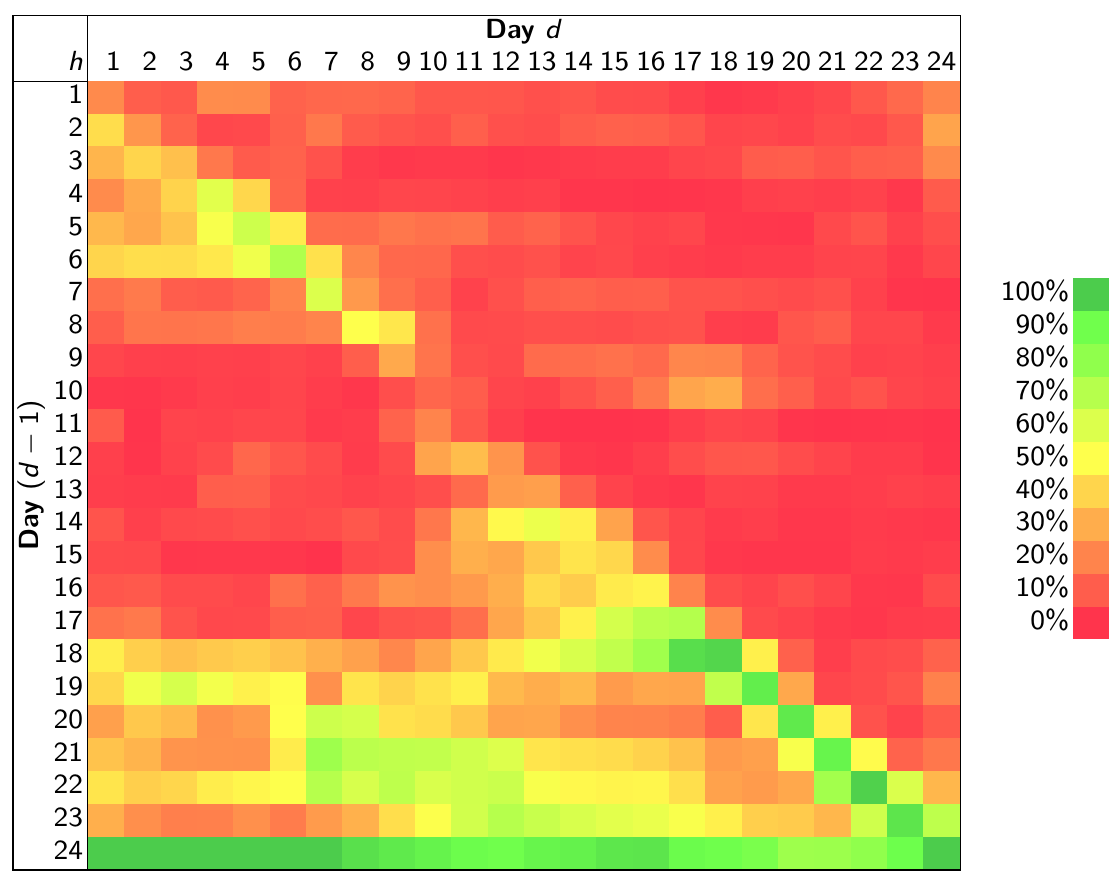}
\caption{Mean occurrence of non-zero $\beta_{h,i}$'s across datasets from 12 power markets. Columns represent the predicted hours on day $d$ and rows the first 24 variables, i.e., $\sum_{i=1}^{24} \beta_{h,i} P_{d-1,i}$ in Eqn.\ \eqref{eq:LEAR}, of a LEAR-type model considered by \cite{zie:wer:18}. A heat map is used to indicate more ($\rightarrow$ green) and less ($\rightarrow$ red) commonly-selected variables.}
\label{fig:param:multi:1}
\end{figure}

This is visualized in Figure \ref{fig:param:multi:1} for the first 24 variables, i.e., $\sum_{i=1}^{24} \beta_{h,i} P_{d-1,i}$, of a LEAR-type model considered by \cite{zie:wer:18} and across datasets from 12 power markets.\footnote{BELPEX price for Belgium, EPEX prices for Switzerland, Germany--Austria and France, EXAA price for Germany--Austria, GEFCom2014 competition data, Nord Pool prices for West Denmark, East Denmark and the system price, OMIE prices for Spain and Portugal, and OTE price for the Czech Republic. The GEFCom2014 dataset covers a 3-year period \cite[2011--2013; see][]{hon:pin:fan:etal:16}, the remaining datasets a 6-year period \cite[July 2010 -- July 2016; see][]{zie:wer:18}.} 
The yellow-green diagonal indicates that the price for hour $h$ on day $d-1$ is a good predictor of the price for the same hour on day $d$. The yellow-green bottom rows were a surprising finding at the time \cite{zie:wer:18} published their paper. They simply mean that late evening prices for day $d-1$ and particularly the last known price, i.e., $P_{d-1,24}$, are good predictors for all hours of the next day. Since then, terms like $\beta_4 P_{d-1,24}$ in Eqn.\ \eqref{eqn:expert} have been added to expert models. 
Interestingly, the performance of LEAR-type models can be further improved by deseazonalizing the data with respect to the long-term seasonal component (LTSC) before estimation \citep{jed:mar:wer:21}, just like in the case of parsimonious regression \citep{now:wer:16} and neural network models  \citep{mar:uni:wer:20IJF}.

\subsection{Deep Learning and the DNN Model}
\label{ssec:DNN}

Starting in the mid 2010s, the EPF research shifted towards models with a larger number of inputs and automatic feature engineering, like the LEAR  described in Section \ref{ssec:LEAR}, and architectures that employ \textit{deep learning} (DL) to obtain better hidden data representations. 
Both families of models are examples of a recent trend called \textit{data-centric} ML, where emphasis is not put on the model, but on input data quality and consistency. Both families use SL/ML methods as means to increase the number of (potential) input features and to reduce the need for human interaction during feature engineering and data processing. The difference is that the second family uses deep architectures, e.g., neural networks with more than one hidden layer \cite[see][for an excellent introduction to DL]{goo:ben:cou:16}.

Deep learning EPF models can be traced back to \cite{wan:zha:che:17}, who proposed an architecture built on stacked \textit{denoising autoencoders} that take a partially corrupted (or noisy) input and are trained to recover the original undistorted input. The DL models that followed were primarily based on the MPL with features modeled as hyperparameters. The most prominent example is probably the DNN model of \cite{lag:rid:sch:18} that has been shown to improve upon parameter-rich linear regression models estimated via the LASSO. 

\begin{figure}[tb]
	\centering
    \includegraphics[height=5cm]{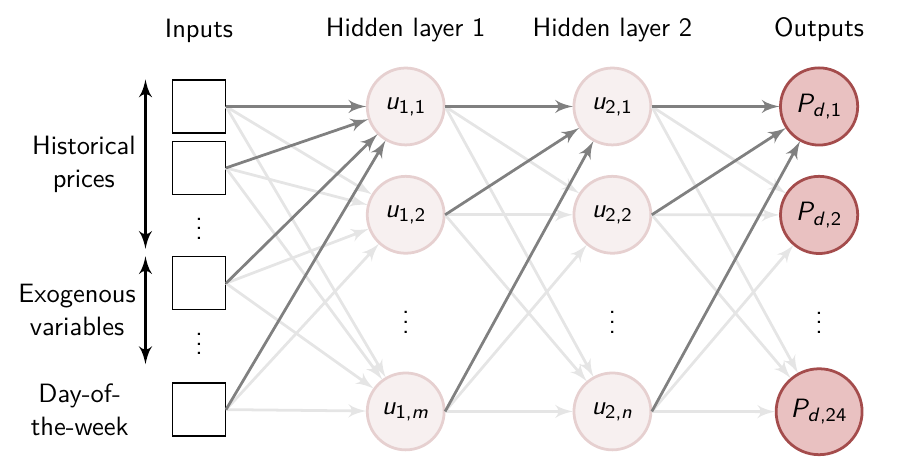}
    \caption{Visualization of the DNN model of \cite{lag:rid:sch:18}, i.e., a feed-forward neural network with two hidden layers and 24 outputs. Like in Figure \ref{fig:ARX:NN}, white squares represent the inputs, $u_{1,1},...,u_{1,m},u_{2,1},...,u_{2,n}$ the hidden nodes, and arrows indicate the flow of information. 
    } 
	\label{fig:DNN}
\end{figure}

The DNN is a feed-forward network with two hidden layers of $m$ and $n$ nodes, and 24 outputs, i.e., it jointly predicts 24 hourly prices $P_{d,1},...,P_{d,24}$, see Fig.\ \ref{fig:DNN}. Its hyperparameters and input features are optimized using the tree-structured Parzen estimator \citep{ber:etal:11}. This is achieved by modeling the features as hyperparameters, with each hyperparameter representing a binary variable that selects whether or not a specific feature is included in the model. Other hyperparameters include the number of neurons per layer, the activation function, the dropout rate, the learning rate, etc. In practice, the model structure can be quite large, \cite{lag:rid:sch:18} report the optimal values in their study of the Belgian market to be $m=239$ and $n=162$.

Given the optimal hyperparameters and features, the DNN is recalibrated on a daily basis to provide next day's electricity price forecasts. Although not strictly mandatory, periodic (e.g., monthly) recalibration of features and hyperparameters can be beneficial. The starting set of input features is the same as for the LEAR model in Eqn.\ \eqref{eq:LEAR}, with the only difference that, for the sake of simplicity, the day-of-the-week is modeled with a multi-valued variable, not a set of 7 dummies \citep{lag:mar:sch:wer:21}. The open-source Python codes for the DNN (and the LEAR) model are available from GitHub (\url{https://epftoolbox.readthedocs.io}).

\subsection{Interpretability and the NBEATSx Model}
\label{ssec:NBEATSx}

The architecture of the \textit{neural basis expansion analysis for time series} (NBEATS) model introduced by \cite{ore:etal:20:nbeats} has the ability to structurally decompose signals making the outputs easily interpretable. A feature whose absence has made it difficult to apply neural networks in many contexts.
Moreover, it has demonstrated state-of-the-art performance on multiple large-scale datasets, including those used in the M4 competition \citep{mak:spi:ass:20}, and it is computationally efficient exhibiting a linear cost with respect to the input size. Recently, it has been successfully applied in mid-term electricity load forecasting  \citep{ore:dud:pel:tur:21}.

\begin{figure}[tb]
	\centering
    \includegraphics[width=.8\textwidth]{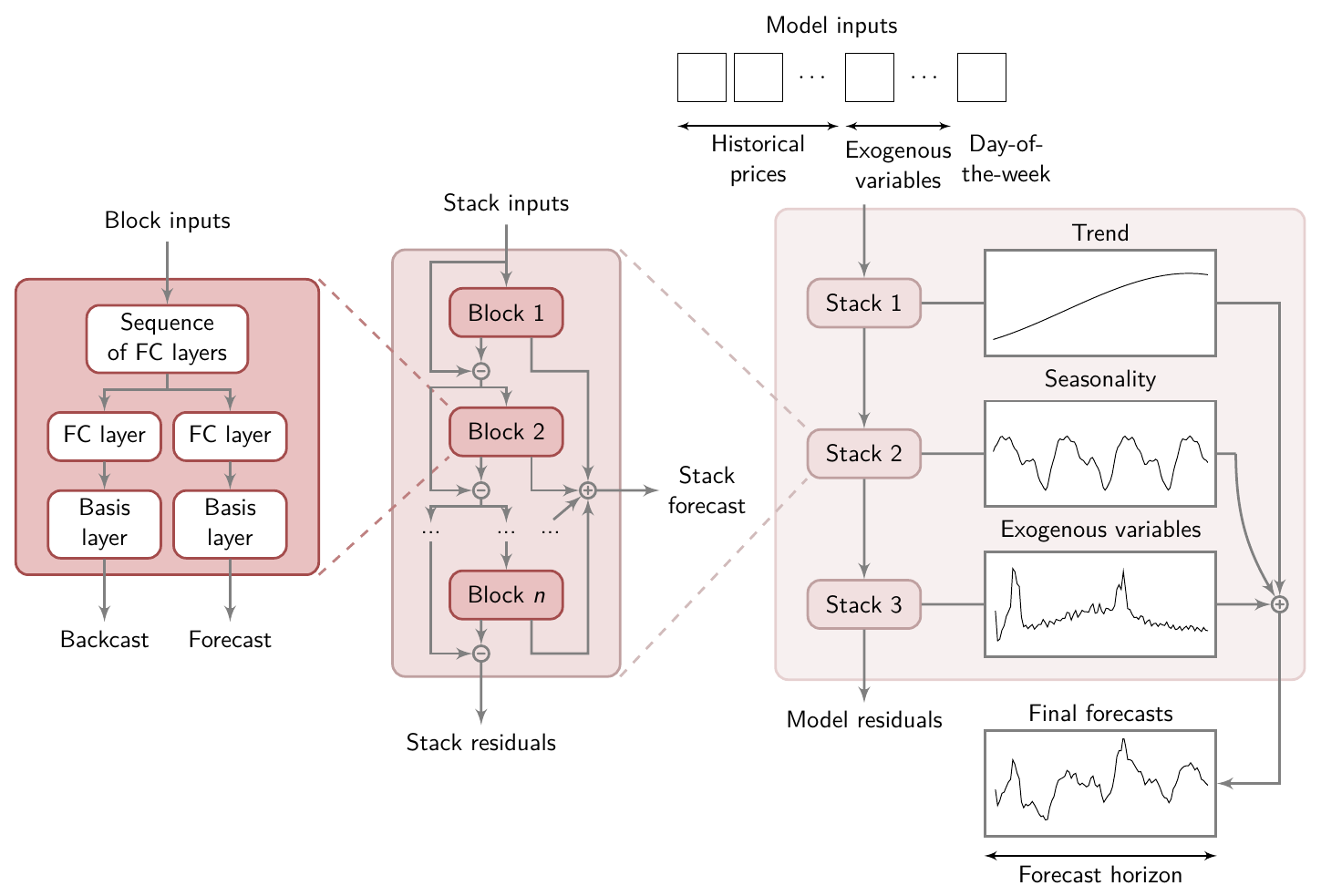}
    \caption{Visualization of the NBEATSx model of \cite{oli:cha:mar:wer:dub:22}. 
    } 
	\label{fig:NBEATSx}
\end{figure}

In general, the decomposition in the NBEATS model is performed by projecting the objective time series onto basis functions in the fundamental blocks of the network structure. Each fundamental block (the dark red rectangles labeled ``block 1'', ..., ``block $n$'' in Fig.\ \ref{fig:NBEATSx}) consists of two parts: (i) a sequence of fully-connected layers (FC) ended with a fork that returns estimated backward and forward expansion coefficients, and (ii) the backward and forward basis layers that map these coefficients via the basis functions onto two block outputs called the backcast and the forecast. The former is the best estimate of the block inputs given the functional space used in the considered block, whereas the latter is the partial prediction that contributes to the final forecast.

The blocks are lined up so that the backcast of each is removed from its inputs, and the residuals are passed to the following block as new inputs. Such a residual recursion is performed consecutively over all blocks in the network. The block forecasts, on the other hand, are summed up to produce the final prediction. The NBEATS architecture groups blocks into stacks that specialize in different types of basis functions. Separate stacks can account for the trend and seasonality by modelling these functions as polynomials and harmonic functions, respectively. Consequently, the final forecasts can be decomposed into interpretable components returned by individual stacks.

The NBEATSx model introduced by \cite{oli:cha:mar:wer:dub:22} adds to this structure a stack (light red rectangle labeled ``stack 3'' in Fig.\ \ref{fig:NBEATSx}) that performs the projection onto exogenous variables. Such an exogenous stack helps to predict the effects induced by holidays and fundamentals (like electric load or RES generation forecasts), and is crucial for EPF. 
While \citeauthor{oli:cha:mar:wer:dub:22} report no significant differences between the NBEATS model and the \textit{exponential smoothing recurrent neural network} (ESRNN) of \cite{smy:20} that has excelled in the M4 competition, the NBEATSx architecture improves over NBEATS by nearly 20\% and up to 5\% over the LEAR (Sec. \ref{ssec:LEAR}) and DNN (Sec. \ref{ssec:DNN}) models. 

The hyperparameters and the input features are optimized in the same way as for the DNN model. However, compared to the DNN, the hyperparameter list also includes: the type and the number of stacks, the number of blocks per stack, the degree of trend polynomials, and the number of Fourier bases. The optimization algorithm also selects the best-performing order of stacks. Open-source Python codes are available from PythonRepo
(\url{https://pythonrepo.com/repo/cchallu-nbeatsx-python-deep-learning}).

\section{Trend \#3: From Statistical to Economic Evaluation}
\label{sec:Trend3}

Over the years a number of authors have criticized the exclusive use of statistical error measures to evaluate and compare forecasts. However, a standardized test ground/procedure for evaluating the economic impact of predictions has not been developed, not only in EPF \citep{hon:pin:etal:20}, but in forecasting in general \citep{pet:etal:22}. And this, despite the fact that already three decades ago \cite{mur:93} postulated that the ``goodness'' of a forecast can be assessed in terms of consistency, quality, and value. 

While \textit{quality} can be readily quantified by commonly used error metrics, see Section \ref{ssec:Error:measures}, the other two characteristics require an explanation. As \cite{mur:93} defines it, \textit{consistency} refers to the correspondence between forecasters' internal, i.e., recorded only in the forecaster's mind, judgments and their forecasts. Since such judgments are, by definition, unavailable to others, consistency cannot be assessed directly. 
Yet, some authors explicitly mention using expert knowledge  to ex-post correct the results from a statistical or a ML model. For instance, \cite{mac:now:16} `manually' expanded or tightened the PIs in their top performing GEFCom2014 competition approach.

The third characteristic, i.e., \textit{value}, refers to the (incremental) economic and/or other benefits to decision makers from using the predictions. For instance, it may reflect additional revenue resulting from improved forecasts or reduced  uncertainty as measured by revenue volatility.
As \cite{yar:pet:21} argue, it is a construct that not only incorporates considerations of the utility to the forecaster, which is discussed in Section \ref{ssec:Economic:Measures} below, but also the computational and opportunity costs. While numerous papers report them, the computational costs are rarely used to compare different methods. One of a few exceptions is an article by \cite{nik:pet:18}, who study the trade-off between optimal versus suboptimal (but less costly) solutions and find that choosing the latter does not necessarily reduce forecast accuracy. Finally, the opportunity costs reflect the resources wasted on implementing a complex method that eventually is not used, because the decision-makers do not have confidence in a model they do  not understand \citep{gre:arm:15}. Yet, both cited papers do not concern EPF. Hence, in Section \ref{ssec:Error:measures} we will briefly review statistical error metrics, emphasizing their pros and cons, and then -- in Section \ref{ssec:Economic:Measures} -- the approaches to evaluating forecast utility, i.e. the main ingredient of the forecast value.

\subsection{Statistical Error Measures}
\label{ssec:Error:measures}

\subsubsection{Point Forecasts}
The most commonly used error metrics for point forecasts include the \textit{mean absolute error} (MAE) and the \textit{root mean squared error} (RMSE), typically across all $H=24$ hours (48 half-hours or 96 quarter-hours) in the test period:
\begin{equation}
    \text{MAE} = \frac{1}{DH} \sum_{d=1}^{D} \sum_{h=1}^{H} |\widehat{\varepsilon}_{d,h}|, \qquad
    \text{RMSE} = \sqrt{\frac{1}{DH} \sum_{d=1}^{D} \sum_{h=1}^{H} \widehat{\varepsilon}_{d,h}^{~2}},
    \label{eq:MAE:RMSE}
\end{equation}
where $\widehat\varepsilon_{d,h} = P_{d,h} - \widehat P_{d,h}$ and $D$ is the number of days in the test period. It is advised to report both absolute and squared errors, especially if regression and neural network models are compared. The reason is that regression-type models are typically estimated using OLS or its variants, as in Eqn.\ \eqref{eq:LASSO}, while NNs are often trained by minimizing absolute errors \citep{lag:rid:sch:18,smy:20,oli:cha:mar:wer:dub:22}.

Both MAE and RMSE are scale-dependent and hence hard to compare across different datasets. The often used in other forecasting contexts \textit{mean absolute percentage error} (MAPE) and its ``symmetric'' variant \cite[sMAPE; see, e.g.,][]{mak:spi:ass:20} are sensitive to values close to zero and may lead to absurd results in EPF. \cite{hyn:koe:06} advocate using the \textit{mean absolute scaled error} (MASE) which is simply the MAE in Eqn.\ \eqref{eq:MAE:RMSE} scaled by the in-sample MAE of a naive\footnote{E.g., a random walk forecast. Note that for seasonal time series of period $\tau$, the time lag should be equal to $\tau$. For instance, in EPF it is common to take $\widehat P_{d,h}^{naive} = P_{d-7,h}$.} forecast. However, the MASE is not recommended for comparisons of models using different calibration windows, since for each model it will be based on a different scaling factor. Instead, \cite{lag:mar:sch:wer:21} recommend using relative measures. For instance, the \textit{relative} MAE (rMAE) which normalizes the MAE by the out-of-sample (not in-sample) MAE of a naive forecast. 

The significance of differences in EPF accuracy is usually evaluated using the \cite{die:mar:95} test for (\textit{unconditional}) \textit{predictive ability} or its generalization -- the \cite{gia:whi:06} test for \textit{conditional predictive ability}. Both tests can be used for nested and non-nested models, as long as the calibration window does not grow with the sample size, but only the latter accounts for parameter estimation uncertainty. However, energy forecasters are not restricted to these two tests, there is a plethora of available approaches \cite[for a review see, e.g., Section 2.12.6 in][]{pet:etal:22}.

The Diebold-Mariano (DM) test is an asymptotic \textit{z}-test of the hypothesis that the mean of the loss differential series is zero. It is based upon the observation that the DM statistic:
\begin{equation}
\mathrm{DM} = \sqrt{DH}\frac{\hat\mu}{\hat\sigma},
\end{equation}
is asymptotically standard normal under the assumption of covariance stationarity of the \textit{loss differential} series: 
\begin{equation}\label{eqn:DM}
\Delta_{d,h} = L_{1,d,h} - L_{2,d,h},
\end{equation}
where $L_{i,d,h}$ is the \textit{score} or \textit{loss function} of model $i$ for day $d$ and load period (e.g., hour) $h$, while $\hat\mu$ and $\hat\sigma$ are respectively the sample mean and standard deviation of $\Delta_{d,h}$. Covariance stationarity may not be satisfied by forecasts in day-ahead electricity markets, since the $H$ predictions for the next day are made at the same time, using the same information set. Hence, either $H$ independent tests \cite[one for each load period of the day;][]{bor:bun:lis:nan:13,now:rav:tru:wer:14,uni:now:wer:16,lag:rid:sch:18,gia:rav:ros:20} or a multivariate variant proposed by \cite{zie:wer:18} are performed \citep{uni:wer:zie:18,hub:mar:wer:19,mar:uni:wer:19,mac:nit:wer:21,oze:yil:21}. The latter jointly tests forecasting accuracy across all $H$ load periods using the `daily' or `multivariate' loss differential series:
\begin{equation}\label{eq:CPA_eq}
    \Delta_{d} = ||\varepsilon_{1,d}||_p - ||\varepsilon_{2,d}||_p,
\end{equation}
where $\varepsilon_{i,d}$ is the $H$-dimensional vector of prediction errors of model $i$ for day $d$,  $||\varepsilon_{i,d}||_p = (\sum_{h=1}^{H} |\varepsilon_{i,d,h}|^p)^{1/p}$ is the $p$-th norm of that vector, with $p=1$ for absolute or $2$ for squared losses.

Like in the DM test, also in the Giacomini-White (GW) test the object of interest is the loss differential series -- univariate or multivariate. We test the null $H_0:\boldsymbol{\phi}=0$ in the following regression (here in the multivariate variant):   
\begin{equation}\label{eqn:GW}
\Delta_d=\boldsymbol{\phi}' {X}_{d-1} + \epsilon_d, 
\end{equation}
where ${X}_{d-1}$ contains elements from the information set on day $d-1$, i.e., a constant and lags of $\Delta_{d}$, and $\epsilon_d$ is an error term. Notice that $\epsilon_d \in R$ is not the 24-dimensional vector $\varepsilon_{i,d}$ of prediction errors from Eq.\ \eqref{eq:CPA_eq}. Sample applications of the GW test in the context of EPF include \cite{mar:ser:wer:18}, \cite{lag:mar:sch:wer:21} and \cite{oli:cha:mar:wer:dub:22}.

\subsubsection{Probabilistic Forecasts}
\label{sssec:PEPF}

While defining error measures for point predictions is relatively straightforward, for probabilistic ones this becomes tricky. The problem is that we cannot observe the true price distribution $F_{P}$, only a single draw from it, i.e., the observed price $P_{d,h}$. Therefore, evaluation of probabilistic forecasts relies on  so-called scoring rules and the notions of reliability, sharpness and resolution.
A \textit{scoring rule} -- also, as in Eq.\ \eqref{eqn:DM}, called score or loss function -- assigns a numerical score $S(\widehat F_P,P_{d,h})$ based on the predictive distribution $\widehat F_P$ and the observed price. A scoring rule is (\textit{strictly}) \textit{proper} if it is (uniquely) optimized in expectation by the true distribution \citep{gne:raf:07}. 
\textit{Reliability} (also called \textit{calibration} or \textit{unbiasedness}) refers to the statistical consistency between $\widehat F_P$ and $P_{d,h}$. For instance, a 95\% prediction interval (PI) is reliable if it covers exactly 95\% of the observed prices. \textit{Sharpness} refers to how concentrated is  $\widehat F_P$. 
Finally, \textit{resolution} refers to how much the predicted density varies over time. Since sharpness and resolution are equivalent when probabilistic forecasts have perfect reliability, evaluating probabilistic predictions boils down to ``maximizing sharpness subject to reliability'' \citep{gne:kat:14,now:wer:18}.


The most intuitive approach to formally check the reliability of a  prediction interval to compute the empirical coverage based on the indicator series of `hits and misses' defined as: $I_{d,h} = 1$ if $P_{d,h} \in$ PI and zero otherwise. EPF studies typically report the empirical coverage itself (\textit{PI coverage probability}, PICP) or the \textit{average coverage error}: ACE $=$ PICP $-$ PINC, where PINC $=\alpha$ is the \textit{PI nominal coverage}. To formally check whether $\mathbb{P}(I_{d,h} = 1) = \alpha$, i.e., the so-called \textit{unconditional coverage} (UC), we can use the \cite{kup:95} test, which verifies whether $I_{d,h}$ is i.i.d.\ Bernoulli with mean $\alpha$. 
Since the latter cannot distinguish between randomly distributed and clustered PI exceedances, \cite{chr:98} introduced the \emph{independence} and \emph{conditional coverage} (CC) tests. The former is tested against a first-order Markov alternative and the latter is a joint test for independence and UC; note, that both can be run for lags larger than one \citep{ber:chr:pel:11}. 
In a continuous setting, i.e., when testing $\widehat F_P$, not just selected PIs, the most common approach is to use the \textit{Probability Integral Transform}:
\begin{equation}\label{eqn:PIT}
\mbox{PIT}_{d,h} = \widehat{F}_{P}(P_{d,h}), 
\end{equation}
which is independent and uniformly distributed if the distributional forecast is perfect. The PIT can be assessed visually \citep{now:wer:18} or formally evaluated using the approach of \cite{ber:01}, which jointly tests for independence and normality, i.e., for conditional coverage. 


Unlike reliability, sharpness is a property of the forecasts only -- the narrower the PI or the more concentrated the predictive distribution the better. Consequently, the PI width itself is a good measure of sharpness. A more elaborate approach relies on proper
scoring rules, which actually assess reliability and sharpness simultaneously \citep{gne:kat:14}. Among them, arguably the most popular is the \textit{pinball} loss, also known as the \textit{linlin}, \textit{bilinear} or \textit{newsboy} loss \citep{ell:tim:16} and has become popular in EPF after the Global Energy Forecasting (GEFCom2014) competition \citep{dud:16,hon:pin:fan:etal:16,mac:now:16}. It is defined by:
\begin{equation}
\label{eq:Pinball}
\mbox{pinball}^{\alpha}
= 
\begin{cases}
(1-\alpha) \left(\widehat{P}^{\alpha}_{d,h} - P_{d,h}\right), & \mbox{for } P_{d,h}<\widehat{P}^{\alpha}_{d,h},\\
\alpha \left(P_{d,h} - \widehat{P}^{\alpha}_{d,h}\right),   & \mbox{for } P_{d,h}\geq\widehat{P}^{\alpha}_{d,h},
\end{cases}
\end{equation} 
where $\widehat{P}^{\alpha}_{d,h}$ is the $\alpha$th quantile of the predictive distribution for day $d$ and load period (e.g., hour) $h$; note, that the pinball score is the function minimized in quantile regression, see Section \ref{ssec:QRA}.
The pinball can be averaged across different quantiles, e.g., 99 percentiles, and across load periods of the target day, e.g., 24 hours, to provide the \textit{aggregate pinball score} (APS). If the grid of quantiles is arbitrarily dense, then the average converges to the \textit{Continuous Ranked Probability Score} \citep{gne:raf:07}:
\begin{align}
\mbox{CRPS}(\widehat{F}_P, P_{d,h}) = \underbrace{\mathbb{E} | \widehat{P}_{d,h} - P_{d,h} |}_{\text{reliability}} - \underbrace{\frac{1}{2} \mathbb{E} | \widehat{P}_{d,h} - \widehat{P}_{d,h}^* |}_{\text{lack of sharpness}},
\label{eq_crps} 
\end{align}
where random variables $\widehat{P}_{d,h}$ and $\widehat{P}^*_{d,h}$ are two independent $\widehat{F}_P$-distributed copies. Probabilistic forecasts can be tested for equal predictive performance using the DM and GW tests, just like point forecasts. In this case $L_{i,d,h}$ is replaced by $S_i(\widehat F_P,P_{d,h})$ in Eq.\ \eqref{eqn:DM}. For sample EPF applications see, e.g., \cite{ser:uni:wer:19}, \cite{abr:bun:20}, \cite{mar:uni:wer:20IJF}, \cite{mun:zie:20} and \cite{uni:wer:21}.

\subsubsection{Path Forecasts}

Compared to evaluating point or probabilistic predictions, evaluating \textit{path} (also called  \textit{ensemble}) forecasts constitutes a challenge -- it requires utilizing scoring rules for multivariate distributions \citep{sch:ham:15}. The commonly used \emph{Dawid-Sebastiani} and \emph{variogram scores} are not strictly proper in the multivariate setting, while the \emph{log-score} requires forecasts of a multivariate density, which may be not available. Hence, the recommended option is the \emph{energy score} proposed by \cite{gne:raf:07}, which is a generalization of the pinball and CRPS scores: 
\begin{equation}
  ES_{d,h} = \underbrace{\frac{1}{M} \sum^M_{i=1} ||P_{d,h}^i-P_{d,h}||_2}_{\text{distance from the prices}} - \underbrace{\frac{1}{2}\frac{1}{M^2} \sum^M_{i=1} \sum^M_{j=1} ||P_{d,h}^i-P_{d,h}^j||_2}_{\text{distance between paths}},  
\end{equation}
where $P^i_{d,h}$ for $i = 1, \ldots, M$ is the $i$-th price path forecast and $||\cdot||_2$ is the Euclidean norm. When minimizing the energy score we are interested in minimizing the average distance between the simulated paths and the actual price trajectory and at the same time maximizing the average distance between the paths. Its use in EPF is limited, though, probably due to the much higher complexity of the problem \citep{mun:zie:20,nar:zie:20b}.

\subsection{Economic Measures}
\label{ssec:Economic:Measures}

There are only a handful of papers which examine the economic impact of EPF errors in a more systematic manner. Interestingly, most of these studies have been published in engineering, not economic or financial journals. The likely reason is that at least a basic knowledge is needed of how power markets, loads and generating units operate. Moreover, as mentioned earlier, there is no standardized test ground/procedure for evaluating the economic impact. Nearly every EPF study considers a different setup. We list them here with the hope of shedding light on this important, but underdeveloped topic.

\subsubsection{Supply- and Demand-Side Perspectives}
\label{sssec:supply:demand}

In one of the earlier studies, \cite{del:vdb:dha:10} take the supply-side point of view and quantify the \emph{profit loss} that can be expected in a price based unit commitment problem, when incorrect price forecasts are used. Simulations reveal that a combined cycle gas turbine (CCGT) is much more sensitive to EPF errors (the profit can easily lie 20\% below the optimal level for a perfect price forecast) than a classic coal fired unit (profit loss rarely exceeds 10\%). More interestingly, negatively biased forecasts (i.e., that predict prices lower than actual) typically yield much higher losses than positively biased predictions.

On the other hand, \cite{zar:can:bha:10} take the demand-side perspective and consider short-term operation scheduling of two typical loads (a process industry owning on-site generation facilities and a municipal water plant with load-shifting capabilities). They introduce the \emph{forecast inaccuracy economic impact} index: $\mbox{FIEI} = [\mbox{cost}(\widehat{P}) - \mbox{cost}(P)] / \mbox{cost}(\widehat{P})$, so that a positive value of FIEI indicates the percentage of the actual cost of buying electricity attributable to EPF errors. The authors report that a 1\% improvement in the MAPE in forecasting accuracy would result in about 0.1\%–0.35\% cost reductions from short-term EPF, but also conclude that the MAPE is not a good measure. 

An interesting concept is considered by \cite{doo:amj:zar:17}, who compute the \emph{financial loss/gain} (FLG) time series, defined as the difference between expected profit of a generator and the actual one. Then, based on the day-ahead forecasts of the FLG series, they propose a bidding strategy. However, by doing so, they do not work with the actual profits but with (another) estimate. 

\cite{mac:nit:wer:19,mac:nit:wer:21} take the perspective of a small RES utility (e.g., with one wind turbine) which has to decide where to sell 1 MW of electricity during each hour of the next day -- in the day-ahead (DA) or the intraday (ID) market. Conditional on the decision, summarized by the \textit{decision variable} based on price forecasts: 
\begin{equation}\label{eq:Y_def}
Y_{d,h} = \left\lbrace
	\begin{aligned}
    & 1\quad \text{if}\ \widehat{P}_{d,h}^{DA}>\widehat{P}_{d,h}^{ID}, \\
    & 0\quad \text{if}\ \widehat{P}_{d,h}^{DA} \leq \widehat{P}_{d,h}^{ID},
    \end{aligned}
    \right.
\end{equation}
they compute the additional income over the benchmark, i.e., selling the production in the DA market, as:
\begin{equation}\label{eq:profit}
    \pi_{d,h}=Y_{d,h}P_{d,h}^{DA} + (1-Y_{d,h})P_{d,h}^{ID}-P_{d,h}^{DA},
\end{equation}
where $P_{d,h}^{DA}$ and $P_{d,h}^{ID}$ are the electricity prices in the DA and ID markets, respectively. While \cite{mac:nit:wer:19} utilize the load forecasts published by the German and Polish system operators, \cite{mac:nit:wer:21} additionally improve the load forecasts for Germany by applying ARX-type models. In both papers, they measure the gains from EPF as the sum of profits in the test period, $\pi=\sum_{d=1}^{D}\sum_{h=1}^{24}\pi_{d,h}$, and conclude that the statistical measures of forecast accuracy -- such as the percent of correct sign classifications of the price spread between the DA and ID markets -- do not necessarily coincide with economic benefits.

\subsubsection{Trading Strategies}

\cite{uni:wer:zie:18} take a trading perspective (different from the supply- or demand-side point of views in Section \ref{sssec:supply:demand}) and consider a naive spot-futures trading strategy in the German market. With a perfect day-ahead forecast the buyer could always choose the lower of the two -- the day-ahead price (unknown when submitting bids) or the futures price. Since this can never be achieved in reality, the authors bias (or perturb) the `crystal-ball' forecast and show that a 0.20 EUR/MWh decrease in the MAE from using one model instead of another would result in ca.\ 90,000 EUR profits, for a 1~GW baseload in 2016.

\cite{chi:zam:zar:pal:18} propose a trading strategy applicable in Ontario’s real-time electricity market. The energy storage operator maximizes profits with optimal scheduling. The schedule is set before the trading period begins, based on the available price forecasts and then it is updated at the end of each hour with a newer price forecasts. 
The authors conclude that such a strategy yields higher profits when using predictions generated by the proposed ARX model with features selected via the Mutual Information technique  \citep{amj:key:zar:11} -- 62\% of the potential saving for `crystal ball' predictions, compared with a number of other EPF approaches, e.g., using the so-called Pre-Dispatch Prices (PDPs; publicly available price predictions published by the system operator IESO) -- 43\% of the potential saving.

\cite{kat:zie:18} propose a multivariate elastic net model (see Section \ref{ssec:LEAR}) for forecasting German quarter-hourly electricity prices. They demonstrate that the ``sell in the high and buy in the low market'' strategy performs well, leading to substantial benefits for both a net buyer and a net seller. On the other hand, the mean-variance approach does not bring economic benefits, but yields an optimal portfolio in terms of the \textit{Sharpe ratio}: 
\begin{equation}\label{eqn:Sharpe}
    SR = \frac{\bar{\pi}}{\sigma},
\end{equation}
where $\bar{\pi}$ denotes the average level of an additional revenue (i.e., $\bar{\pi} = \pi/24D$; see also Eq.\ \eqref{eq:profit}) and $\sigma$ is the standard deviation of the time series of revenues. As such, the Sharpe ratio can be used to assess the trade-off between revenue and uncertainty. However, there are more performance measures \citep{eli:sch:07,aue:15}, including measures based on drawdowns (e.g., Calmar ratio, Sterling ratio), based on partial moments (e.g., omega ratio, Sortino ratio) and based on the Value-at-Risk (VaR; e.g., excess return on VaR, conditional Sharpe ratio). Whether they will turn out to be useful in the EPF context remains yet to be checked.

\begin{figure}[tbp]
	\centering 
	\includegraphics[width = 0.45 \linewidth]{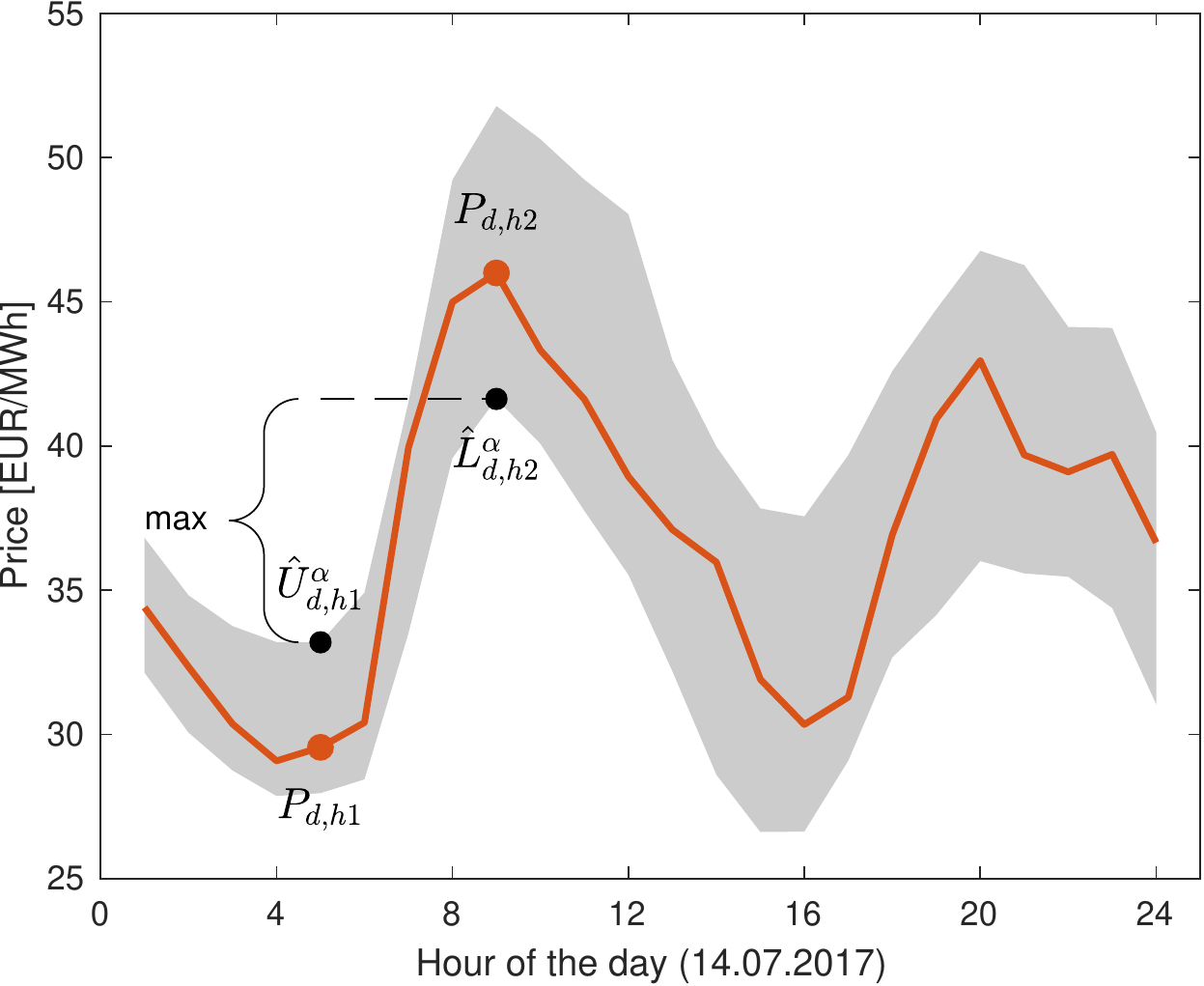}~~~~
	\includegraphics[width = 0.45 \linewidth]{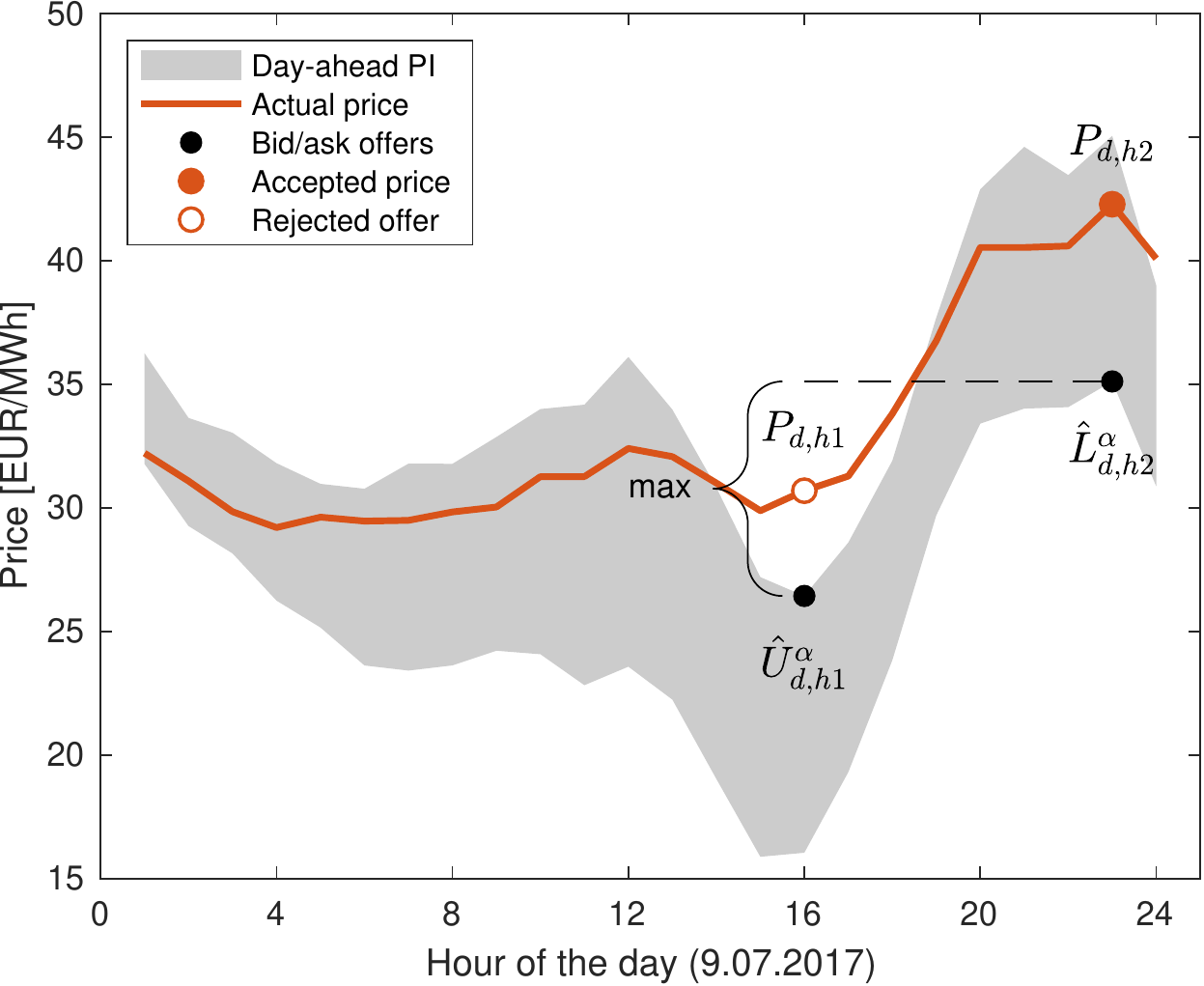}
	\caption{Illustration of the trading strategy considered by \cite{uni:wer:21} using German EPEX data for Friday, July 14th (\textit{left panel}) and Sunday, July 9th (\textit{right panel}) 2017. The day-ahead forecast of the PI is plotted in  gray, the bids for the selected hours are indicated with black dots and the actual price trajectory is in orange. Note, that on July 9th the buy order is not accepted because $P_{d,h1}>\hat{U}^{\alpha}_{d,h1}$.
	}
	\label{fig:stategy}
\end{figure}

\cite{uni:wer:21} propose a strategy for market participants having access to storage capacity. They consider a realistic setup, inspired by the Virtual Power Plant analyzed in \cite{sik:etal:19}, in which the company owns a 1.25 MW battery with an efficiency of 80\% per charge and discharge cycle, that cannot be discharged below 20\% of the nominal capacity (i.e., 0.25 MW) due to technical limitations. The strategy is straightforward: each day buy energy and charge the battery when the price is low (generally in the early morning hours) and discharge and sell when the price is high (generally in the afternoon hours). 
Using probabilistic forecasts of the DA prices in the Polish market, the authors determine both the time (buy $h1$ and sell $h2$ hours) and the prices of limit orders submitted to the power exchange. They formulate and solve the following maximization problem:
\begin{equation}
\max_{\{h1, h2\}} \left( 0.8~\hat L^{\alpha}_{d,h2} - \hat U^{\alpha}_{d,h1} \right) \qquad \text{subject to} \qquad h1 < h2.
\end{equation}
The optimizer selects the lowest price of a given day based on the upper quantile forecast $\hat U^{\alpha}_{d,h1}$ and the highest price based on the lower quantile forecast $\hat L^{\alpha}_{d,h2}$. The company then submits the bid to buy 1 MW for $\hat U^{\alpha}_{d,h1}$ at hour $h1$ and simultaneously the offer to sell 0.8 MW at $\hat L^{\alpha}_{d,h2}$ at hour $h2$; two sample solutions for the German EPEX market are depicted in Fig.\ \ref{fig:stategy}. If both offers are accepted in the day-ahead market, as in the left panel of Fig.\ \ref{fig:stategy}, the profit for a given day equals $0.8P_{d,h2} - P_{d,h1}$. However, the probability of each offer to be accepted in the market is equal to $\frac{1-\alpha}{2}$. If one of them is rejected, as in the right panel of Fig.\ \ref{fig:stategy} for hour $h1$, the energy has to be bought or sold in the balancing market.

This strategy is further modified in \cite{uni:wer:22}. Now, all trading is exclusively executed in the day-ahead market. To do so, a twice larger energy storage capacity (2.5 MW) is required to trade the same volume (0.8-1 MW). The idea is to always remain in an intermediate state of the battery, for which both charging and discharging 1 MW is possible. When the bid or the ask is rejected, the authors propose to close the position on the next day by submitting a market order (i.e., with no price limit). They conclude that the Smoothing Quantile Regression Averaging (SQRA) approach they propose outperforms the benchmarks in terms of statistical error metrics (Kupiec test, GW test for the pinball score) in all four considered markets (German EPEX, Scandinavian Nord Pool, Iberian OMIE, North American PJM). However, when the trading strategy is executed, SQRA forecasts lead to higher profits only in two markets (EPEX, PJM). The authors hypothesize that the poor performance for NP and OMIE is due to a twice lower average intraday price spread, i.e., the gap between the maximum and the minimum hourly price for a given day.

\section{Further Reading}

The first review, published when electricity markets were still in their infancy:
\begin{itemize}
    \item Bunn, D., \citeyear{bun:00}. Forecasting loads and prices in competitive power markets. Proceedings of the IEEE, 88(2), 163-169.
\end{itemize}
The first comprehensive EPF review, postulating the need for objective comparative studies and speculating on the future research directions:
\begin{itemize}
    \item Weron, R., \citeyear{wer:14}. Electricity price forecasting: A review of the state-of-the-art with a look into the future. International Journal of Forecasting 30, 1030-1081. [Open access]
\end{itemize}
Two thorough treatments of probabilistic EPF, presenting much needed guidelines for the rigorous use of methods, measures and tests, in line with the paradigm of maximizing sharpness subject to reliability:
\begin{itemize}
    \item Nowotarski, J., Weron, R., \citeyear{now:wer:18}. Recent advances in electricity price forecasting: A review of probabilistic forecasting.  Renewable and Sustainable Energy Reviews 81, 1548-1568.
    \item Ziel, F., Steinert, R., \citeyear{zie:ste:18}.   Probabilistic mid- and long-term electricity price forecasting. Renewable and Sustainable Energy Reviews 94, 251-266.
\end{itemize}
A review of energy (load, price, wind and solar generation) forecasting, with a discussion of two challenging problems that deserve rigorous investigation -- close-loop forecasting and (economic) valuation of forecasts:
\begin{itemize}
    \item Hong, T., Pinson, P., Wang, Y., Weron, R., Yang, D., Zareipour, H., \citeyear{hon:pin:etal:20}. Energy forecasting:  A review and outlook. IEEE Open Access Journal of Power and Energy 7, 376-388. [Open access]
\end{itemize}
A recent review with a set of guidelines/best practices for EPF, introducing the \texttt{epftoolbox}\footnote{Freely available for download from: \url{https://epftoolbox.readthedocs.io/en/latest}.} with Python codes for two highly competitive benchmark models (LEAR, DNN):
\begin{itemize}
    \item Lago, J., Marcjasz, G., De Schutter, B., Weron, R., \citeyear{lag:mar:sch:wer:21}. Forecasting day-ahead electricity prices: A review of state-of-the-art algorithms, best practices and an open-access benchmark.  Applied Energy 293, 116983. [Open access]
\end{itemize}
A popular science article on the evolution of machine learning models in EPF:
\begin{itemize}
    \item J\c{e}drzejewski,  A.,  Lago,  J.,  Marcjasz,  G.,  Weron,  R.,  \citeyear{jed:lag:mar:wer:22}.  Electricity price forecasting: The dawn of machine learning. IEEE Power \& Energy Magazine 20(3), 24-31.
    
\end{itemize}

\bibliographystyle{elsarticle-harv}
\addcontentsline{toc}{section}{References}
\setlength{\bibsep}{0pt plus 0.3ex}
\small
\bibliography{epf}

\end{document}